%% file: main.tex
\documentclass[sigconf]{acmart}
\AtBeginDocument{%
  }

\copyrightyear{2026}
\acmYear{2026}
\setcopyright{cc}
\setcctype{by}
\acmConference[Koli Calling '26]{26th Koli Calling International Conference on Computing Education Research}{November 05--08, 2026}{Koli, Finland}
\acmBooktitle{26th Koli Calling International Conference on Computing Education Research (Koli Calling '26), November 05--08, 2026, Koli, Finland}
\acmDOI{10.1145/3856208.3856234}
\acmISBN{979-8-4007-2752-8/2026/11}

\begin{document}

\title{It's Complicated: Factors Influencing the Effectiveness of LLM-Generated Feedback in a Live Programming Course}

\author{Tsvetomila Mihaylova}
\affiliation{\institution{Aalto University}\city{Espoo}\country{Finland}}
\email{tsvetomila.mihaylova@aalto.fi}

\author{Arto Hellas}
\affiliation{\institution{Aalto University}\city{Espoo}\country{Finland}}
\email{arto.hellas@aalto.fi}

\author{Bita Akram}
\affiliation{\institution{North Carolina State University}\city{Raleigh}\state{NC}\country{United States}}
\email{bakram@ncsu.edu}

\author{Narges Norouzi}
\affiliation{\institution{University of California, Berkeley}\city{Berkeley}\state{CA}\country{United States}}
\email{norouzi@berkeley.edu}

\author{Peter Brusilovsky}
\affiliation{\institution{University of Pittsburgh}\city{Pittsburgh}\state{PA}\country{United States}}
\email{peterb@pitt.edu}

\author{Juho Leinonen}
\affiliation{\institution{Aalto University}\city{Espoo}\country{Finland}}
\email{juho.2.leinonen@aalto.fi}

\renewcommand{\shortauthors}{Mihaylova et al.}

\input{00-abstract}





\maketitle

\input{00-custom}

\input{10-introduction}

\input{20-background}

\input{30-methods}

\input{40-results}

\input{50-discussion}

\input{55-limitations}

\input{60-conclusions}

\begin{acks}
This work was supported by Research Council of Finland grants \#367787 and \#356114, and by the National Science Foundation grants \#2426837, \#2426838, and \#2426839.
\end{acks}

\balance
\bibliography{main}
\bibliographystyle{abbrv}

\end{document}

%% file: 00-abstract.tex
\begin{abstract}
We report on three iterations of an LLM-generated feedback system deployed in a live web development course, spanning three underlying models and three feedback formats. We ask how feedback presence and its characteristics (length, complexity, format) relate to student performance, whether prior feedback exposure is associated with final-project performance, and how performance changes after first exposure. Feedback was associated with more favorable immediate outcomes in several comparisons, but the pattern varied by outcome and period. In pooled models, longer feedback was associated with lower next-submission correctness, more remaining attempts, and longer resubmission time. Prior same-part feedback was associated with lower final-project solve probability. Student-level comparisons found differences on final-project steps begun before exposure, while within-student changes across steps were uncertain. These comparisons do not identify the causal effect of feedback. Changes on standard exercises were more favorable among exposed students, but the groups' different course-stage splits limit causal interpretation. Bounded message length and student choice of format are directions for experimental evaluation. Our final-project analysis captures follow-up performance rather than directly measuring learning or understanding.
\end{abstract}

%% file: 00-custom.tex
\newcommand{\dbbeforefeedback}{\texttt{Before\_Feedback}}
\newcommand{\dbafterfeedback}{\texttt{After\_Feedback}}

\newcommand{\diagnosticfeedback}{\texttt{Diagnostic Feedback}}
\newcommand{\singleissuefeedback}{\texttt{Single-Issue Feedback}}
\newcommand{\generalfeedback}{\texttt{General Feedback}}
\newcommand{\nofeedback}{\texttt{No Feedback}}

\newcommand{\novicestudents}{\texttt{Novice Students}}
\newcommand{\advancedstudents}{\texttt{Advanced Students}}

\newcommand{\votegood}{\texttt{Good}}
\newcommand{\votebad}{\texttt{Poor}}

\newcommand{\periodone}{\texttt{Period 1}}
\newcommand{\periodtwo}{\texttt{Period 2}}
\newcommand{\periodthree}{\texttt{Period 3}}

\newcommand{\coursepartoldcontent}{\texttt{Old Content}}
\newcommand{\coursepartnewcontent}{\texttt{New Content}}

\newcommand{\assignmentcompletion}{\texttt{Assignment Completion}}
\newcommand{\timetosuccess}{\texttt{Time to Success}}
\newcommand{\attemptstosuccess}{\texttt{Attempts to Success}}

\newcommand{\ie}{\textit{i.e.}}
\newcommand{\ex}{\textit{ex.}}


%% file: 10-introduction.tex
\section{Introduction}

There is a growing body of research on LLM-generated feedback for programming exercises \cite{leinonen2023using,RN5981,mihaylova-etal-2026-effects}, but moving from a research prototype to a production course is not trivial. A comparison of models responding to beginner programmers' help requests found differences in issue identification and instruction following, including false positives and solution-like responses despite requests not to provide them \cite{hellas2023help}. Feedback design therefore needs evaluation in its actual model and task context. Prior work has compared guidance formats and immediate task performance \cite{mihaylova-etal-2026-effects}; this study adds sequential deployments, joint analyses of feedback length, reading complexity and format, and temporally ordered final-project and within-student comparisons.

We deployed three feedback formats in an online introductory web development course, varying the number of issues identified and the guidance provided. The format analyses focus on Single-Issue and Diagnostic Feedback, which were available in all three periods; General Feedback was offered only in the first. The no-LLM-feedback comparison retains the platform's ordinary grader feedback. We address four research questions:

\begin{itemize}
    \item[\textbf{RQ1:}] How is receiving LLM-generated feedback associated with student performance, compared with receiving ordinary grader feedback alone?
    \item[\textbf{RQ2:}] Among submissions that received feedback, which characteristics of that feedback — length, complexity, format — are associated with better subsequent performance?
    \item[\textbf{RQ3:}] How is prior LLM-generated feedback associated with performance on final-project exercises without direct feedback?
    \item[\textbf{RQ4:}] How does performance change after students first receive feedback, compared with changes among students who never receive it?
\end{itemize}

We find that feedback is associated with more favorable immediate outcomes in several comparisons, but the pattern varies by outcome and period. Longer messages are associated with lower next-submission correctness, more remaining attempts, and longer resubmission time in pooled models. Prior same-part feedback is associated with lower final-project solve probability. Complementary student-level timing comparisons show differences on final-project steps begun before exposure and uncertain within-student changes, while standard-exercise improvement differs between exposed and never-exposed students. The different exposure definitions, outcomes, and course-stage splits prevent these comparisons from establishing a causal explanation.

Based on these results, we discuss two directions for evaluation: bounding feedback length and offering students a choice between a single-issue hint and a fuller diagnostic account. These proposals require experimental evaluation.

%% file: 20-background.tex
\section{Background}

Automated programming-feedback tools commonly report test outcomes, compiler diagnostics, and static-analysis findings \cite{messer2024grading}. Such messages can be difficult for beginners to interpret \cite{becker2016effective}. A systematic review of automated feedback generation identifies limitations in the support provided to students, difficulties adapting tools to teachers' needs, and substantial variation in how effectiveness is evaluated \cite{10.1145/3231711}. Evidence on enhancing compiler messages is mixed: Becker reported improvements in compiler-error outcomes \cite{becker2016effective}, whereas Denny et al. found no significant benefit from the syntax-error enhancements they evaluated \cite{denny2014syntax}. These results motivate evaluating particular message designs rather than assuming that additional explanation is necessarily beneficial.

Readability is one aspect of message quality. Denny et al. identified message length, jargon, sentence structure, and vocabulary as factors associated with novices' judgments of programming-error-message readability \cite{denny2021readability}. These judgments are distinct from whether a message enables a correct repair. Researcher evaluations have judged some LLM-generated error explanations more interpretable and actionable than the original programming messages \cite{leinonen2023using}. An instructor evaluation of Llama 3 feedback found that ratings varied across prompting strategies and task difficulty, with different configurations performing best on the two tasks examined \cite{RN6198}. Evaluations also identify limitations: Xiao et al. report cases of inaccurate or inappropriate guidance, while showing that more concrete hint levels can sometimes help novices when abstract guidance is insufficient \cite{xiao2024exploring}.

Studies of task performance provide a complementary perspective. In a study of error resolution, fine-tuned explanations reduced time to fix, while differences in reruns and task failures were associated with the debugging task rather than message style \cite{11176024}. A prior comparison of General, Single-Issue, and Diagnostic Feedback found that all three formats were associated with faster solving than the no-LLM-feedback baseline \cite{mihaylova-etal-2026-effects}. The estimates differed by format, but did not establish a general ordering in which less guidance produced better outcomes. Together, these findings motivate examining length, reading complexity, and format alongside subsequent performance.

Classroom deployments examine how students use feedback as well as the responses that systems produce. CodeAid was deployed over a 12-week semester in a programming course of roughly 700 students, offering conceptual explanations, pseudocode, and feedback intended to avoid directly supplying solutions \cite{kazemitabaar2024codeaid}. Its findings emphasize transparency, student control, and the trade-off between immediate assistance and cognitive engagement. In a small study of programming students' chatbot interactions, help-seeking and elaboration were common, while critical thinking, organization, and effort regulation were less evident \cite{RN6178}.

Help-seeking is also associated with students' motivation. An exploratory study linked mastery-approach orientation to more conceptual and fewer implementation-focused requests to an LLM-based chatbot \cite{10.1145/3769994.3770009}. In a study offering planning, debugging, and optimization hints, planning-hint use was associated with higher problem-solving performance than requesting no hints; the overall comparisons for the other hint types were uncertain \cite{RN6189}. Students' interpretations of the feedback they receive provide another perspective on its use. In a mixed-methods study using eye tracking and think-aloud data, expressed understanding of a feedback component was associated with subsequent code improvement, whereas disagreement or a stated need for more information was associated with seeking another component \cite{10.1145/3769994.3770034}.

Students' reported experiences need to be distinguished from performance measured through tasks or assessments. A survey-based analysis reported a positive association between LLM-use frequency and students' perceived impact on their programming performance \cite{10.1145/3768421.3768464}. Comparative studies, however, have not consistently demonstrated learning or performance advantages. An exploratory randomized comparison involving 33 students detected no significant differences in learning gains, completion time, or code accuracy between a context-aware tutor, a general-purpose chatbot, and no AI support \cite{10.1145/3769994.3770025}. A quasi-experimental course deployment found favorable perceptions of the feedback tools but did not establish an improvement in assessed performance; the authors considered the passing-rate evidence inconclusive \cite{Martinez-Araneda2025}.

A further question is whether benefits on supported tasks are accompanied by better performance on later tasks without the same assistance. This distinction predates LLM-based feedback. In short block-based tasks with novice crowdworkers, Marwan et al. found that code hints with explanations improved supported-task performance, while evidence of later benefits was restricted to previously practiced objectives in a post hoc analysis of the self-explanation condition \cite{marwan2019hints}.

For LLM-generated hints, a classroom experiment found that many students rated the hints as useful and that the experimental group performed better on tasks with hints \cite{RN5981}. On eight tasks where GPT-generated hints were unavailable, that group initially had lower submission success but solved tasks faster; the success difference diminished with further submissions. Ordinary platform feedback remained available. These mixed results do not establish a general loss of independent performance.

A controlled study of 69 novice learners aged 10--17 likewise separated AI-supported practice from later assessment without the tool. An AI code generator improved code-authoring performance during practice, while overall correctness comparisons on immediate and one-week delayed assessments were not significant \cite{kazemitabaar2023codegenerators}. The population and code-generation intervention differ from our course setting, but the design illustrates the distinction between supported practice and later independent performance.

Evidence about longer-term outcomes remains limited. A scoping review of GenAI-enabled personalization in computer science education found that most included studies were single-semester deployments, leaving longer-term retention and transfer insufficiently established \cite{ai7010006}. A broader review of GenAI-related harms identified concerns about over-reliance and metacognitive regulation, while distinguishing hypothetical harms, informal observations, and measured evidence \cite{bernstein2025beyond}.

The present study examines feedback presence and message characteristics alongside immediate submission outcomes, final-project performance, and student-level changes around first feedback exposure. The follow-up comparisons extend the behavioral analysis, but do not directly measure learning gains or identify causal transfer.

%% file: 30-methods.tex
\section{Methodology}

\subsection{Context}

We deployed three iterations of an LLM-generated feedback system in an online introductory web development course. The platform returned automated grader feedback, including test results and available compilation or runtime errors. AI-generated hints supplemented this feedback and were labeled as AI-generated. \autoref{tab:feedack-types-per-period} shows the three analysis periods, models used and feedback types. During active feedback availability in \periodone, eligible incorrect submissions triggered random allocation to one of the feedback types or no LLM feedback when the student had no prior submissions to that exercise. For \periodtwo{} and \periodthree{}, opting in enabled automatic \singleissuefeedback{} or \diagnosticfeedback{} following each eligible incorrect submission, without a separate request for each hint. Returned hints were displayed in the grading results. Within each student--exercise pair in these periods, the platform reused a supported feedback format found in recent submissions; otherwise, \singleissuefeedback{} or \diagnosticfeedback{} was selected randomly with equal probability.

The course combines client-side Svelte/SvelteKit and HTML/CSS with server-side Deno/Hono applications, database access, authentication, and automated testing. Feedback exposure denotes a recorded AI-generated hint attached to a submission, with timing based on that submission's timestamp; it does not establish that the message was read. \nofeedback{} denotes the absence of such a hint, while ordinary grader feedback remains available. Use of other assistance, including the separate course AI chat assistant, was not measured. The three formats were intended to provide a guided single-issue hint (\singleissuefeedback{}), a broader diagnostic account with less solution guidance (\diagnosticfeedback{}), or broader guided feedback (\generalfeedback{}). Legacy novice, advanced, and level-not-mentioned labels are mapped to these categories, respectively. Final-project exercises are successive steps of an overarching application spread across course parts. A student can therefore attempt an early project step before first receiving feedback on a later standard exercise.

\begin{table}[htbp]
    \centering
    \caption{Description of the three periods of time in the study. }
    \begin{tabular}{p{2cm}|c|p{4cm}}
    \toprule
        Period & Model & LLM feedback formats offered \\
        \midrule
        \periodone:
        Sep.--Dec. 2025 & o3-mini & \generalfeedback, \singleissuefeedback, \diagnosticfeedback \\
        \periodtwo:
        Jan.--May 2026 & GPT-5-mini & \singleissuefeedback,  \diagnosticfeedback \\
        \periodthree: 
        May 31--Jun. 2026 & GPT-5.5 & \singleissuefeedback,  \diagnosticfeedback \\
         \bottomrule
    \end{tabular}
    \par\smallskip
    {\footnotesize The analyzed no-LLM-feedback groups occur in all three periods and retain ordinary grader feedback. They include observations outside active AI-feedback availability.\par}
    \label{tab:feedack-types-per-period}
\end{table}

The periods are analysis calendar bins: before 1 January 2026, 1 January through 30 May 2026, and from 31 May onward (local course time). The calendar periods include times when AI feedback was unavailable. Size eligibility required a nonempty submission payload of at most 2,000 characters; eligibility alone did not guarantee a hint. We use that eligibility at the first attempt when defining standard-exercise analysis pairs, but retain all subsequent attempts when determining outcomes.

\input{31-data-stats}

\subsection{Modeling Approach}

For RQ1--RQ3, we use mixed-effects models \cite{lindstrom1988newton,brysbaert2018power} with random intercepts for students and exercises to account for repeated observations of each student and exercise.

For time-based outcomes, we use linear mixed-effects models (LMMs) on log-transformed time, fit using \texttt{statsmodels} version 0.14 \cite{seabold2010statsmodels}. For binary outcomes (correctness, solved), we use Bernoulli generalized linear mixed-effects models. For count-based outcomes (attempts), we use negative binomial generalized linear mixed-effects models (NB-GLMMs) with a log link to account for overdispersion. Binary and count models were fit using \texttt{Bambi} version 0.17 \cite{capretto2022bambi}, via Hamiltonian Monte Carlo (NUTS) sampling with 2{,}000 tuning and 2{,}000 sampling draws per chain across four chains, with target acceptance 0.95 and seed 42. RQ1b attempts used 4{,}000 sampling draws per chain. We use Bambi's default priors. Model estimates, within-period
contrasts, and fitting diagnostics are reported in the supplementary materials.

\textbf{Statistical significance}. For all model families, we treat a fixed effect as significant when its interval estimate excludes zero: for LMMs, the 95\% confidence interval reported by \texttt{statsmodels}; for Bayesian Bernoulli and NB-GLMMs, the 94\% highest-density interval (HDI). Supplementary model tables show fixed-effect coefficients whose intervals exclude zero in \textbf{bold}. Mixed-model inference uses intervals rather than $p$-values.

Predictions set student and exercise random intercepts to zero. Binary and count plots show posterior means with 94\% HDIs. Time plots transform the fitted log scale as $\exp(\eta)-1$ and show 95\% confidence intervals; these are conditional median-time predictions under the log-normal model, not arithmetic mean times. Random-effect comparisons report SDs on each model's link scale, rather than percentages of outcome variance.

\subsubsection{General model structure}

Let $y_{ij}$ denote the outcome for student $i$ on exercise $j$, and let $\eta_{ij}$ denote the linear predictor:
\begin{align}
\eta_{ij}
&=
\beta_0
+ \boldsymbol{\beta}^{\top}\mathbf{X}_{ij}
+ u_{0i}
+ v_{0j},
\label{eq:general}
\end{align}
where $\mathbf{X}_{ij}$ contains the fixed effects relevant to the analysis, $u_{0i} \sim \mathcal{N}(0, \sigma_u^2)$ is a student-specific random intercept, and $v_{0j} \sim \mathcal{N}(0, \sigma_v^2)$ is an exercise-specific random intercept.

For elapsed-time outcomes (\emph{Time to Next Submission}, \emph{Time to Solve}), we model:
\begin{align}
\log(1+y_{ij}) &= \eta_{ij} + \varepsilon_{ij},
&
\varepsilon_{ij} &\sim \mathcal{N}(0,\sigma^2).
\label{eq:lmm}
\end{align}

For binary outcomes (\emph{Next-Submission Success}, \emph{Ever Solved}), we model:
\begin{align}
y_{ij} &\sim \text{Bernoulli}(p_{ij}),
&
\text{logit}(p_{ij}) &= \eta_{ij}.
\label{eq:bernoulli}
\end{align}

For count outcomes (\emph{Attempts to Solve}, \emph{Remaining Attempts}), we model:
\begin{align}
y_{ij} &\sim \text{NegBin}(\mu_{ij},\alpha),
&
\log(\mu_{ij}) &= \eta_{ij},
\label{eq:negbin}
\end{align}
where $\alpha > 0$ is the overdispersion parameter.

The analyses below use different populations and fixed effects in $\mathbf{X}_{ij}$.

\subsection{RQ1: Feedback Presence and Standard-Exercise Performance}

\begin{itemize}
    \item RQ1a: Submission level. Following an eligible incorrect first submission, how is feedback presence associated with (i) correctness of the next submission and (ii) elapsed time to that submission?
    \item RQ1b: Student--exercise pair level. Among eligible pairs that begin with an incorrect attempt, how is receiving feedback during the exercise sequence associated with (i) eventual solving, (ii) attempts through first success, and (iii) elapsed time to first success?
\end{itemize}

\subsubsection{RQ1a: Submission-Level}
For each eligible first attempt on a standard exercise that was incorrect, we checked whether it carried \singleissuefeedback{}, \diagnosticfeedback{}, or no LLM feedback. General Feedback was excluded because it occurred only in Period 1. We linked this attempt to the actual next submission on the same exercise before applying eligibility filters, retained observations with a known next outcome within 30 minutes, and modeled next-submission correctness and elapsed time. Both models included categorical period and a feedback-by-period interaction, with student and exercise random intercepts. We estimate feedback-format contrasts within each period by combining the main and interaction coefficients and their joint uncertainty. Formally, with $\textit{FeedbackType}_{ij} \in \{\text{None}, \text{Single-Issue}, \text{Diagnostic}\}$ and $\textit{Period}_{ij} \in \{1, 2, 3\}$:
\begin{align}
\eta_{ij}
&=
\beta_0
+ \beta_1\,\textit{FeedbackType}_{ij}
+ \beta_2\,\textit{Period}_{ij}
\nonumber\\
&\quad
+ \beta_3\,(\textit{FeedbackType}_{ij}\times\textit{Period}_{ij})
+ u_{0i}
+ v_{0j},
\label{eq:rq1a}
\end{align}
with success modeled via Equation~\ref{eq:bernoulli} and time via Equation~\ref{eq:lmm}, each substituting $\eta_{ij}$ from Equation~\ref{eq:rq1a}.

\subsubsection{RQ1b: Student-exercise pair}
For standard-exercise pairs whose first actual attempt was incorrect and feedback-eligible, we classified feedback by the first message received through the first correct submission, or through the observed sequence if never solved. General Feedback pairs were excluded. The solved model includes all remaining pairs, including slow successes and observed failures. The attempts and time models use solved pairs completed within 30 minutes: attempts count all submissions through first success, and time runs from the first attempt to that success. All three models use the linear predictor in Equation~\ref{eq:rq1a}, with solved modeled via Equation~\ref{eq:bernoulli}, attempts via Equation~\ref{eq:negbin}, and time via Equation~\ref{eq:lmm}.

\subsection{RQ2: Feedback Characteristics and Subsequent Performance}
Among feedback-carrying submissions in RQ1a, we modeled next-submission correctness and time, plus the number of subsequent actual submissions until first success for pairs eventually solved. Feedback length is measured in characters. Reading complexity is represented by the Flesch--Kincaid grade level \cite{kincaid1975readability}, computed using \texttt{textstat} version 0.7.13 \cite{textstat0713}. The measure combines sentence length and syllables per word, with higher scores corresponding to a higher approximate US school-grade reading level. It is a textual proxy, not a measure of conceptual or code difficulty. We passed stored feedback strings directly to the scoring function, without first removing code fragments, identifiers, or Markdown. Sentence boundaries followed the package's default punctuation-based rules. Length and grade are standardized on the RQ2 success/time cohort after all exclusions; the same means and SDs are used for prediction. Models contain one common length slope and one common complexity slope, feedback format (Single-Issue as reference), categorical period, their format-by-period interaction, and student/exercise random intercepts. Length plots show conditional Period~1 predictions over each format's observed 5th--95th percentile length range, with complexity at the fitted-cohort mean. Formally, with $\textit{Length}_{ij}$ and $\textit{Complexity}_{ij}$ standardized to zero mean and unit variance:
\begin{align}
\eta_{ij}
&=
\beta_0
+ \beta_1\,\textit{Format}_{ij}
+ \beta_2\,\textit{Period}_{ij}
\nonumber\\
&\quad
+ \beta_3\,(\textit{Format}_{ij}\times\textit{Period}_{ij})
\nonumber\\
&\quad
+ \beta_4\,\textit{Length}_{ij}
+ \beta_5\,\textit{Complexity}_{ij}
+ u_{0i}
+ v_{0j},
\label{eq:rq2}
\end{align}
with success modeled via Equation~\ref{eq:bernoulli}, remaining attempts via Equation~\ref{eq:negbin}, and time via Equation~\ref{eq:lmm}, each substituting $\eta_{ij}$ from Equation~\ref{eq:rq2}.

\subsection{RQ3: Feedback Presence and Final-Project Performance}

\textbf{Method}. Unlike RQ1 and RQ2, RQ3 uses students with substantial course engagement: submissions in all seven parts, at least 70\% of eligible standard exercises attempted, at least four standard-exercise pairs with an incorrect first attempt, and activity in one period only. These criteria indicate engagement, not verified course completion. We restrict this cohort to Period 2, the only period with a usable comparison group (\autoref{tab:rq3_period_balance}). We exclude every final-project pair containing direct LLM feedback. Exposure means that feedback on a standard exercise in the same course part occurred strictly before the first attempt on the final exercise. The solve-rate model includes all eligible final pairs; attempts and time use solved pairs completed within 72 hours, counting attempts only through first success. A zero-second final solve means the first submitted version passed; this submission-log measure does not include development before the first submission. The models have crossed student/exercise random intercepts and a binary prior-exposure predictor, $\textit{SawFeedback}_{ij}\in\{0,1\}$:

\begin{align}
\eta_{ij}
&=
\beta_0
+ \beta_1\,\textit{SawFeedback}_{ij}
+ u_{0i}
+ v_{0j},
\label{eq:rq3}
\end{align}
with solve rate modeled via Equation~\ref{eq:bernoulli}, attempts via Equation~\ref{eq:negbin}, and time via Equation~\ref{eq:lmm}, each substituting $\eta_{ij}$ from Equation~\ref{eq:rq3}.

\subsection{RQ4: Performance Change After Feedback Exposure}

Using the same Period~2 engaged cohort, we split each student's history by the first attempt on each exercise: before versus after their first standard-exercise feedback timestamp, or at the midpoint of their own sequence if never exposed. An episode assigned to the before group can continue after exposure. We summarize solved pairs only, with at least two solved pairs on each side for struggling standard exercises and one on each side for final exercises. Actual zero-second solves are retained; elapsed times are uncapped in this descriptive analysis. Improvement is each student's median before minus median after, separately for attempts through first success and time to that success. A two-sided Mann--Whitney test compares improvement scores between groups, and a Wilcoxon signed-rank test examines paired change within each group. Signed-rank tests use the normal approximation and discard zero differences, consistently across the within-group and complementary paired comparisons.

A complementary final-project analysis summarizes one median per student in each of three timing groups: Before Exposure, After Exposure, and Never Exposed. Comparisons against never-exposed students use Mann--Whitney tests; before versus after uses a paired Wilcoxon test restricted to students observed in both periods. RQ4 examines student-level changes in attempts and elapsed time using student-wide exposure, complementing RQ3's part-local regression analyses. Tests use $p<0.05$ without multiplicity adjustment and are interpreted as exploratory. Non-significant differences do not establish equivalence or absence of a causal effect.

\subsection{Use of Generative AI}

We used OpenAI’s GPT‑6 Astra Pro to critique the manuscript and suggest refinements to the analysis methodology, including analysis inclusion criteria, temporal definitions of feedback exposure, and interpretation of statistical comparisons. The tool also assisted with literature checking and language editing. The authors evaluated the suggestions and made the final methodological and interpretive decisions. 

%% file: 31-data-stats.tex
\subsection{Data Statistics}
\autoref{tab:period_stats} summarizes the prepared data: unique students, submissions, exercises, and student--exercise pairs in each period. Periods 1 and 2 are of comparable scale; Period 3 is smaller. These totals describe the course data before analysis-specific filters.
\input{table-periods}

\autoref{tab:feedback_by_period} reports the feedback composition of the fitted RQ1a cohort. General Feedback is excluded from RQ1--RQ2 models because it is absent from Periods 2 and 3. The initial allocation included a no-LLM-feedback condition, but the analyzed no-feedback groups also contain submissions outside active feedback availability and therefore are not pure randomized control groups. Later no-feedback comparisons additionally include students who did not opt in.
\input{table-composition}

98.7\% of recorded next submissions after eligible first incorrect standard attempts occur within 30 minutes (\autoref{tab:time_interval_stats}). We retain that interval for immediate outcomes and for RQ1b solved-pair attempts/time, while retaining all eligible struggling pairs for eventual solving. The cutoff limits long breaks but does not establish continuous engagement; prior work shows why elapsed gaps require care as time-on-task measures~\cite{edwards2022practical,leinonen2021fine}.

\input{table-intervals}

\autoref{tab:rq3_period_balance} shows the single-period students with substantial engagement. Period 2 provides the most usable exposed/never-exposed comparison. RQ3 and RQ4 use this cohort, with additional outcome-specific restrictions. \autoref{tab:experiment_ns} reports the analysis sample sizes.
\input{table-balance}
\input{table-model-ns}

\begin{figure}[htbp]
\centering
\includegraphics[width=\linewidth]{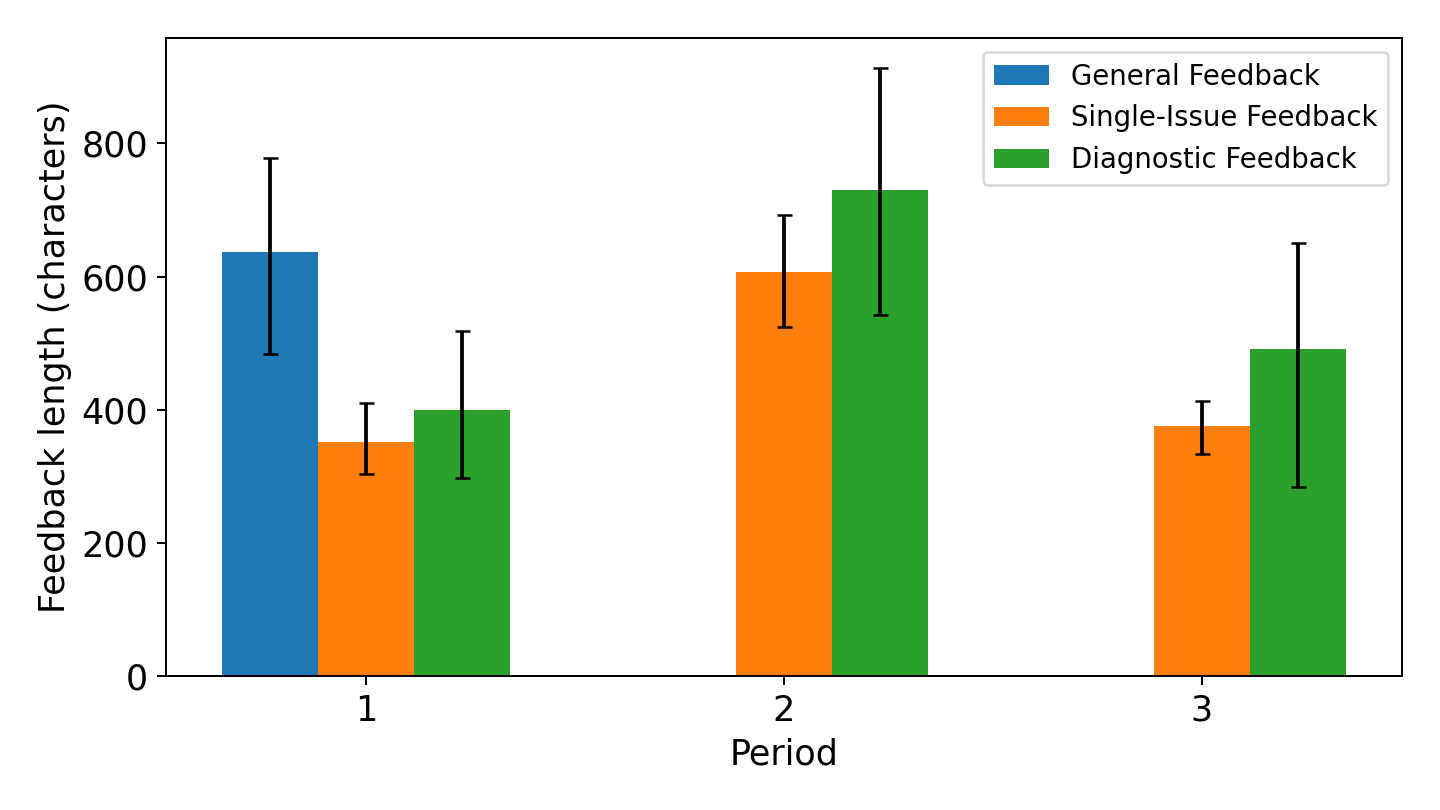}
\caption{Feedback length across all feedback-bearing submissions, by period and format. Bars show medians and error bars show the interquartile range. General Feedback occurs only in Period 1.}
\Description{Grouped bars of median feedback character length by period and format, with interquartile ranges.}
    \label{fig:hint_length}
\end{figure}

\autoref{fig:hint_length} summarizes all feedback-bearing submissions, a broader population than RQ2's first-attempt cohort. Feedback lengths differ by format and deployment period. The RQ2 models therefore include measured character length and reading complexity rather than treating format as a substitute for these properties.

%% file: table-periods.tex
\begin{table}[htbp]
\caption{Per-period scale of the prepared course data. Students and pairs can occur in multiple periods.}
\label{tab:period_stats}
\begin{tabular}{lrrrr}
\toprule
Period & Users & Submissions & Exercises & Pairs \\
\midrule
1 & 419 & 79,438 & 134 & 26,738 \\
2 & 525 & 83,374 & 134 & 29,844 \\
3 & 163 & 10,420 & 134 & 4,778 \\
\bottomrule
\end{tabular}
\end{table}

%% file: table-composition.tex
\begin{table}[htbp]
\caption{Feedback-type composition of the fitted RQ1a cohort; General Feedback is excluded.}
\label{tab:feedback_by_period}
\begin{tabular}{lrrr}
\toprule
Period & No Feedback & Single-Issue & Diagnostic \\
\midrule
1 & 3507 & 1275 & 1324 \\
2 & 6047 & 750 & 861 \\
3 & 818 & 116 & 111 \\
\bottomrule
\end{tabular}
\end{table}

%% file: table-intervals.tex
\begin{table}[htbp]
\caption{Time to the actual next submission after an eligible first incorrect standard-exercise attempt with a known next outcome ($n=16,400$), before the 30-minute and General Feedback exclusions. Interval upper bounds are inclusive. Percentages are rounded independently to one decimal place and may not sum to 100\%; aggregate percentages are calculated from counts.}
\label{tab:time_interval_stats}
\begin{tabular}{lrr}
\toprule
Interval & $n$ & \% \\
\midrule
0--1 min & 8,335 & 50.8 \\
1--5 min & 6,878 & 41.9 \\
5--10 min & 692 & 4.2 \\
10--15 min & 152 & 0.9 \\
15--30 min & 136 & 0.8 \\
30--60 min & 55 & 0.3 \\
$>$60 min & 152 & 0.9 \\
\bottomrule
\end{tabular}
\end{table}

%% file: table-balance.tex
\begin{table}[htbp]
\caption{Students meeting the engagement and single-period criteria, by lifetime logged feedback exposure. RQ3 uses temporally prior, part-local exposure.}
\label{tab:rq3_period_balance}
\begin{tabular}{lrrr}
\toprule
Period & Never exposed & Ever exposed & Total \\
\midrule
1 & 7 & 130 & 137 \\
2 & 54 & 66 & 120 \\
3 & 3 & 2 & 5 \\
\bottomrule
\end{tabular}
\end{table}

%% file: table-model-ns.tex
\begin{table}[htbp]
\caption{Final analysis sizes, after outcome-specific exclusions. RQ1--RQ3 use student--exercise observations; RQ4 uses students.}
\label{tab:experiment_ns}
\begin{tabular}{llr}
\toprule
Analysis & Outcome / unit & $n$ \\
\midrule
RQ1a & Next outcome/time & 14,809 \\
RQ1b & Ever solved & 15,087 \\
RQ1b & Attempts/time, solved & 13,842 \\
RQ2 & Next outcome/time & 4,437 \\
RQ2 & Remaining attempts & 4,381 \\
RQ3 & Ever solved & 2,242 \\
RQ3 & Attempts/time, solved & 2,127 \\
RQ4 & Standard, students & 115 \\
RQ4 & Final, students & 107 \\
\bottomrule
\end{tabular}
\end{table}

%% file: 40-results.tex
\section{Results}

Full model estimates, within-period contrasts, student-level summaries,
fitting diagnostics, and selected feedback prompt templates are
available in the supplementary materials on OSF.\footnote{\url{https://osf.io/4dpf5/overview?view_only=74690991a39c469bae98c36889f99c60}}

\subsection{RQ1: LLM-Generated Feedback Presence and Student Performance}

\input{rq1a}

\input{rq1b}

\input{rq2}

\input{rq3}

\input{rq4}

%% file: rq1a.tex
\subsubsection{RQ1a: Submission-Level}

\textbf{Success}. Receiving \singleissuefeedback{} or \diagnosticfeedback{} was associated with higher odds of getting the next submission correct in \periodone{} (log-odds coefficients 0.254, 94\% HDI [0.109, 0.389], and 0.420 [0.282, 0.568], respectively). \diagnosticfeedback{}'s association was smaller in \periodtwo{} than in \periodone{}: the interaction interval excluded zero. Within-period contrasts supported a positive association for \singleissuefeedback{} in \periodtwo{} and \diagnosticfeedback{} in \periodthree{}, but not for the other format in those periods. \periodthree{} had much less data and wider intervals.

\textbf{Time to next submission}. In \periodone{}, \singleissuefeedback{} was associated with a faster next submission than \nofeedback{}, while \diagnosticfeedback{} was associated with a slower one. In \periodtwo{}, both formats were associated with slower resubmission than \nofeedback{}. In \periodthree{}, the Single-Issue contrast favored faster resubmission, while the Diagnostic contrast was uncertain.

\textbf{Student and exercise variation}. For both outcomes, the estimated exercise random-intercept SD exceeded the student SD (success: 1.033 vs.\ 0.473 on the logit scale; time: 0.450 vs.\ 0.273 on the log-time scale).

\textbf{Overall}. Feedback was associated with higher next-submission success in several format-period comparisons, but the pattern differed across periods and outcomes.

\begin{figure}
    \centering
    \includegraphics[width=0.9\linewidth]{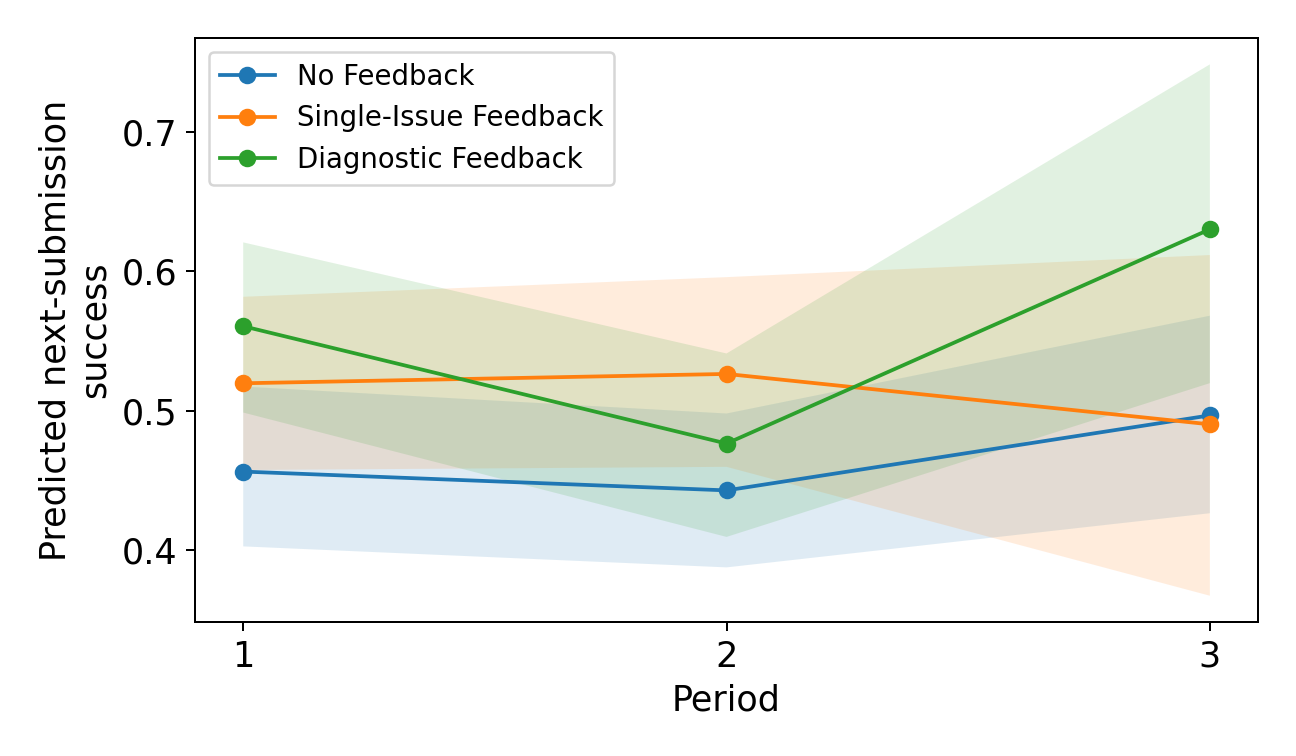}
    \includegraphics[width=0.9\linewidth]{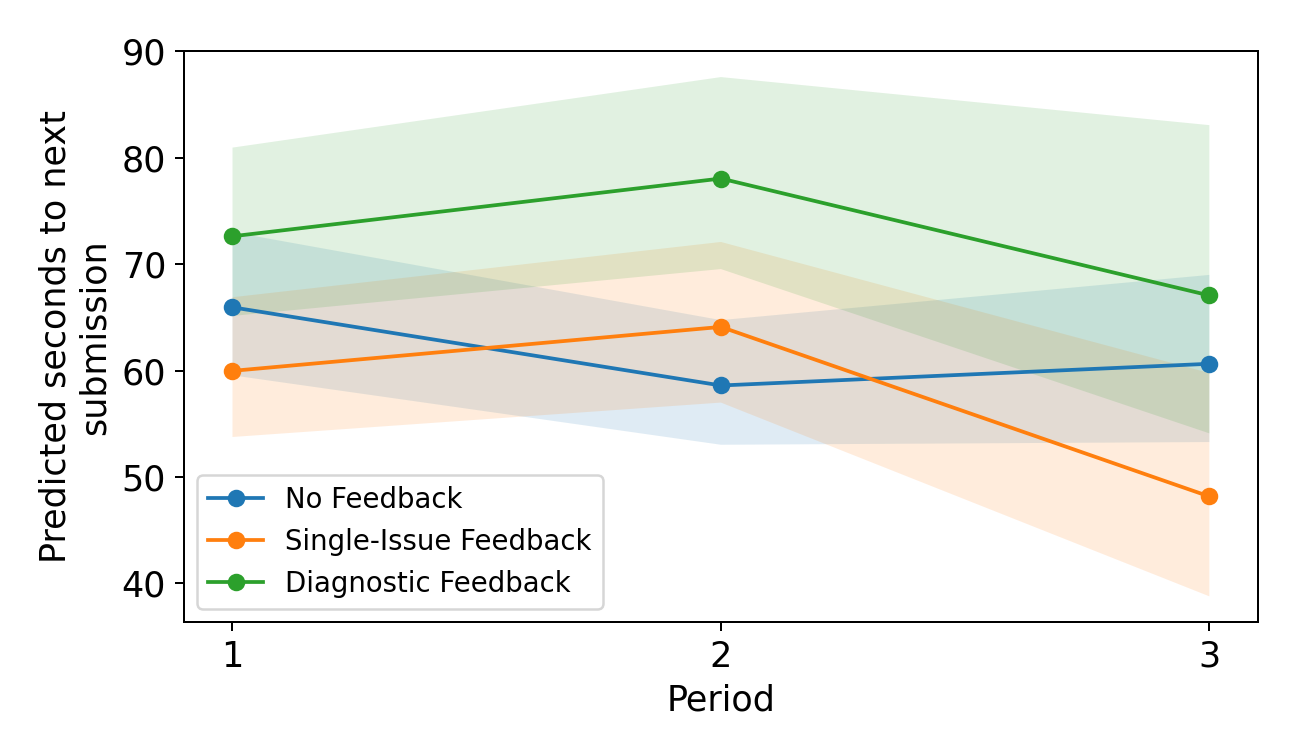}\\
    \caption{RQ1a: Conditional predictions for next-submission success (top) and time to the actual next submission (bottom), by feedback type and period. Student and exercise random effects are set to zero. Success bands are 94\% HDIs; time bands are 95\% CIs for back-transformed log-time predictions. The same 14,809 first-incorrect standard-exercise observations contribute to both models.}
    \Description{Two line plots of conditional next-submission correctness and elapsed time by period and feedback format, with uncertainty bands.}
    \label{fig:rq1a}
\end{figure}

%% file: rq1b.tex
\subsubsection{RQ1b: Student--Exercise Pair Level}

\textbf{Ever solved}. None of the within-period feedback contrasts had a 94\% HDI excluding zero. This model includes all 15,087 eligible struggling pairs, retaining slow successes as well as observed failures. High completion rates leave limited room for differences. Student-level heterogeneity exceeded exercise-level heterogeneity for this outcome (random-intercept SD 3.772 vs.\ 1.414 on the logit scale).

\textbf{Attempts to solve}. Among the 13,842 pairs solved within 30 minutes, both \singleissuefeedback{} and \diagnosticfeedback{} were associated with fewer actual attempts through first success in \periodone{} and \periodtwo{}. The Period~3 contrasts were uncertain.

\textbf{Time to solve}. Both feedback formats were associated with faster solving in \periodone{}. In \periodtwo{}, the Single-Issue contrast was uncertain and Diagnostic Feedback was associated with slower solving than No Feedback. In \periodthree{}, Single-Issue Feedback was associated with faster solving, while the Diagnostic contrast was uncertain.

\textbf{Overall}. Associations with fewer attempts were supported in the first two periods among quickly solved pairs, while timing associations varied. Eventual solving showed no clear feedback association.

\begin{figure}
    \centering
    \includegraphics[width=0.88\linewidth]{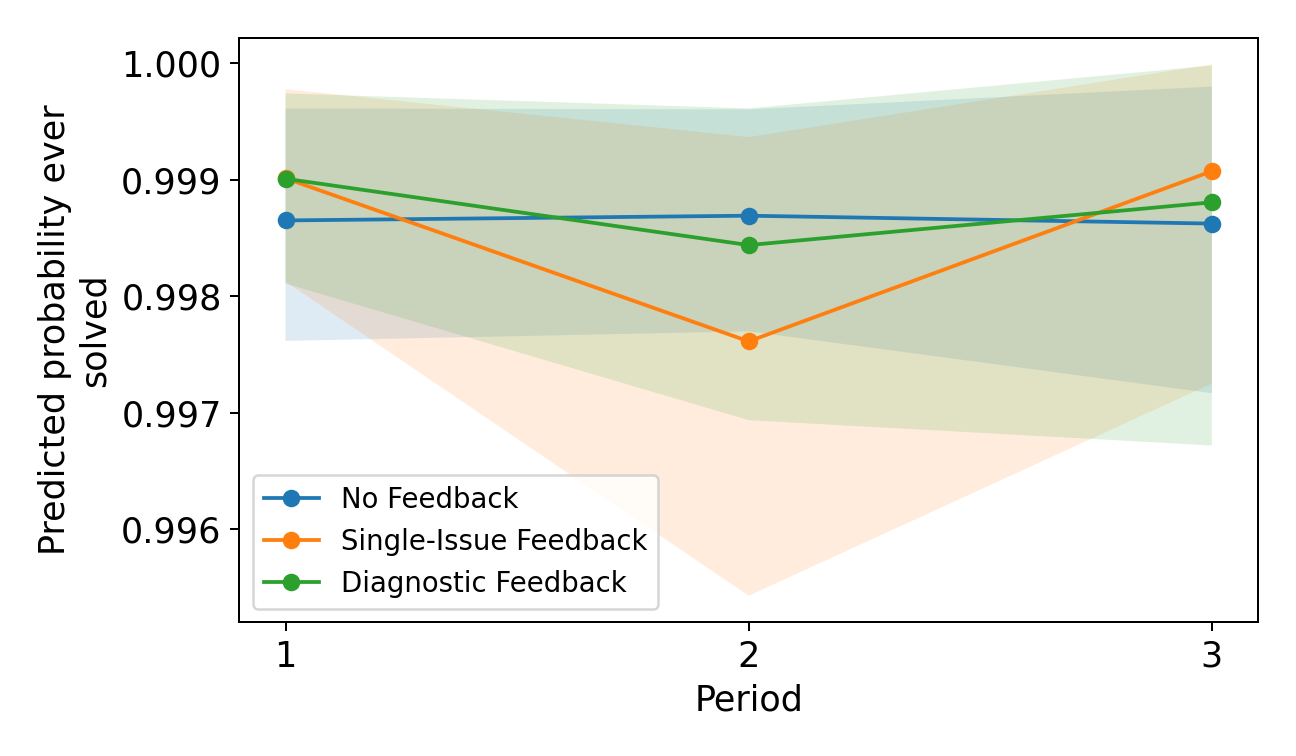}
    \includegraphics[width=0.88\linewidth]{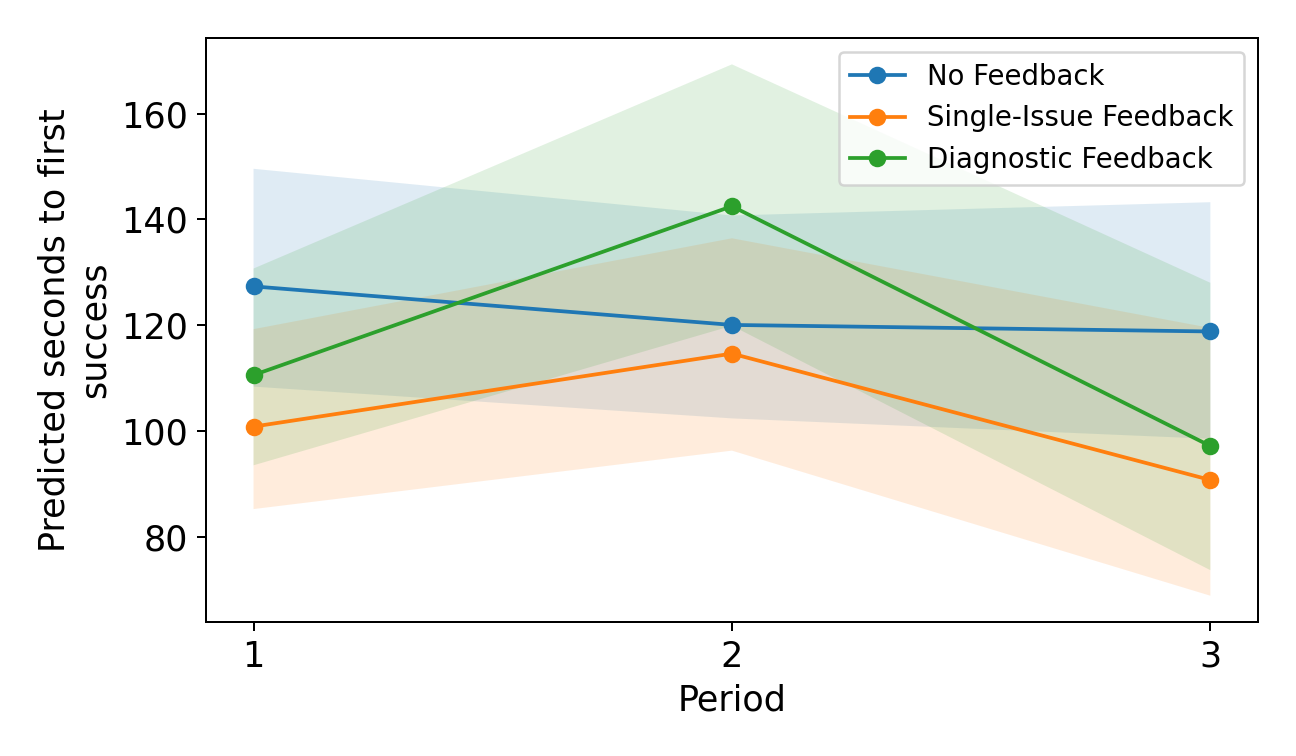}\\
    \includegraphics[width=0.88\linewidth]{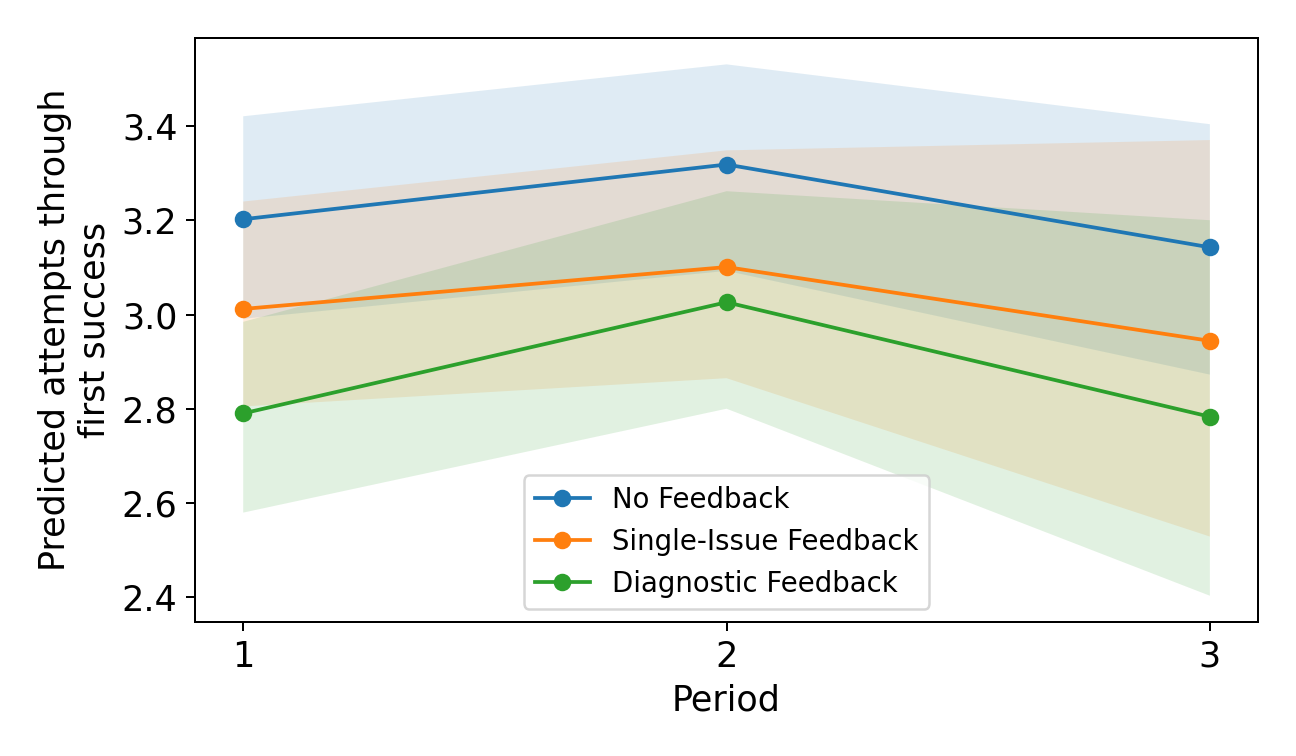}\\
    \caption{RQ1b: Conditional predictions for ever solving (top; $n=15,087$), time to first success (middle), and attempts through first success (bottom; both $n=13,842$). The solved model retains slow successes and failures; time and attempts condition on success within 30 minutes. Student/exercise random effects are zero. Bands are 94\% HDIs for binary/count predictions and 95\% CIs for back-transformed log-time predictions.}
    \Description{Three line plots of conditional eventual solving, elapsed time, and attempts through first success by period and feedback format, with uncertainty bands.}
    \label{fig:rq1b}
\end{figure}

%% file: rq2.tex
\subsection{RQ2: Feedback Characteristics}

\begin{figure}
    \centering
    \includegraphics[width=0.88\linewidth]{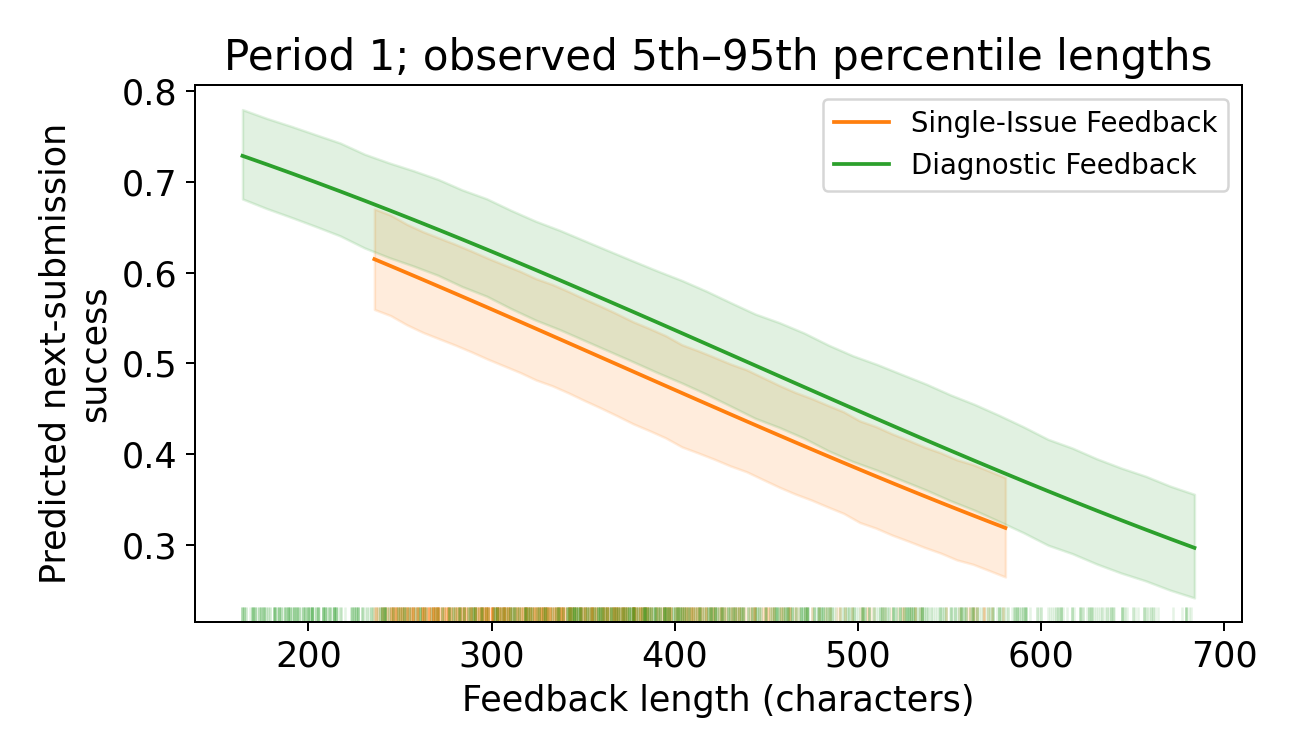}
    \includegraphics[width=0.88\linewidth]{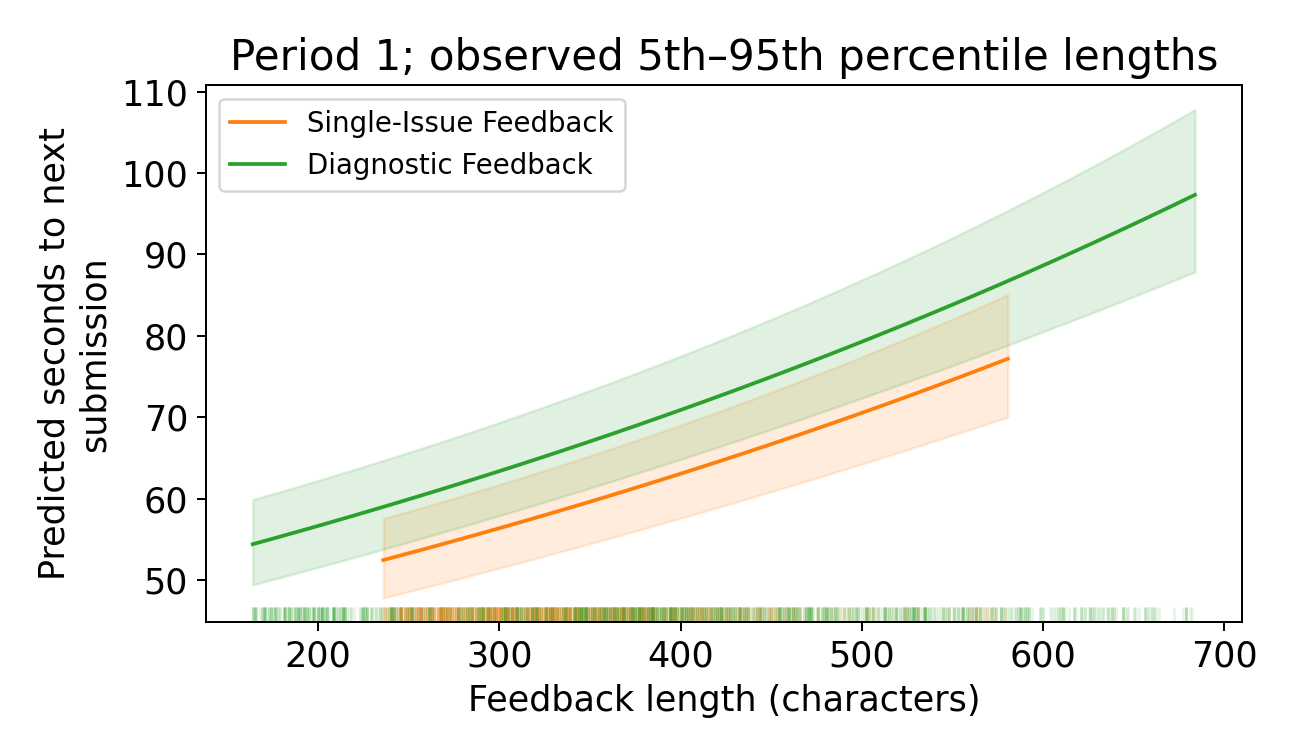}\\
    \includegraphics[width=0.88\linewidth]{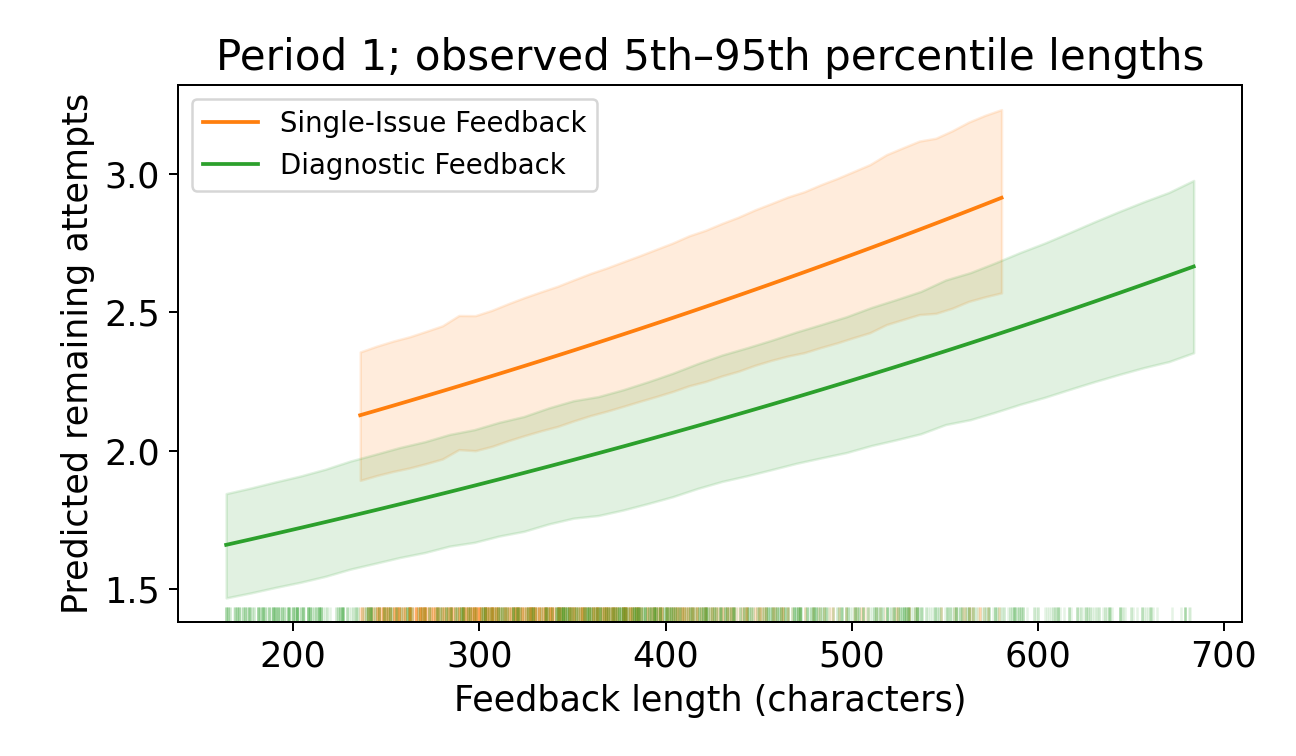}\\
    \caption{RQ2: Conditional predictions for next-submission success (top), time to the next submission (middle), and remaining actual attempts to first success (bottom), as a function of feedback length. Curves use Period 1, mean fitted-cohort complexity, and zero student/exercise effects. Each format is plotted over its observed Period~1 5th--95th percentile length range; rug marks show observations within that range. Bands are 94\% HDIs for correctness/counts and 95\% CIs for back-transformed log-time. Success/time use 4,437 observations; remaining attempts use 4,381 eventually solved pairs.}
    \Description{Three plots of next-submission correctness, elapsed time, and remaining attempts against feedback length, with format-specific observed length ranges, data rugs, and uncertainty bands.}
    \label{fig:rq2}
\end{figure}

\textbf{Feedback length}. Longer feedback was associated with lower odds of next-submission correctness, more remaining attempts, and longer time to the next submission (Figure~\ref{fig:rq2}). Per SD of length (204.9 characters in the fitted cohort), the coefficients were $-0.733$ (94\% HDI [$-0.840$, $-0.631$]) for correctness, 0.187 ([0.153, 0.220]) for remaining attempts, and 0.226 (95\% CI [0.193, 0.259]) for log-time. These are common slopes estimated across periods and formats.

\textbf{Feedback complexity}. Complexity showed a weaker pattern than length. Its interval spanned zero for correctness and remaining attempts. Higher Flesch--Kincaid grade was associated with longer time to next submission (coefficient 0.040 per SD, 95\% CI [0.012, 0.069]).

\textbf{Feedback format}. Holding measured length and complexity constant, \diagnosticfeedback{} was associated with higher next-submission correctness than \singleissuefeedback{} in Periods 1 and 3; the Period~2 contrast was uncertain. Diagnostic Feedback was associated with fewer remaining attempts in all three periods. Diagnostic Feedback was associated with longer resubmission times in Periods 1 and 2; the Period~3 contrast was uncertain. These associations suggest a trade-off between fewer attempts and faster resubmission.

\textbf{Student and exercise variation}. Exercise random-intercept SDs exceeded student SDs in these models (success: 0.882 vs.\ 0.468; attempts: 0.454 vs.\ 0.288; time: 0.331 vs.\ 0.229). The estimates are on their respective link scales.

\textbf{Overall}. Length had a clear pooled association with all three subsequent outcomes. Complexity was less consistently associated, while format differences depended on outcome and period.

%% file: rq3.tex
\subsection{RQ3: Feedback Exposure and Final-Project Performance}

\autoref{tab:rq3-summary} reports conditional predictions and exposure coefficients for final-project solve rate, attempts, and time to solve. Exposure here means standard-exercise feedback in the same course part before the first final attempt. Seven final pairs with direct feedback were excluded.

\textbf{Solve rate}. Prior feedback exposure was associated with lower final-project solve probability (log-odds coefficient $-1.096$, 94\% HDI [$-2.074$, $-0.136$]). Observed solve rates were 95.6\% for prior-exposed pairs (721/754) and 98.1\% for pairs without prior same-part exposure (1,459/1,488). Conditional model predictions were 99.2\% and 99.7\%, respectively, with student and exercise effects set to zero. These predictions differ from observed rates because they describe a particular random-effect setting; the small difference between them does not by itself establish negligible practical importance.

\textbf{Attempts and time to solve}. Neither exposure coefficient had an interval excluding zero among final pairs solved within 72 hours (attempts: 0.013, 94\% HDI [$-0.099$, 0.116]; log-time: $-0.002$, 95\% CI [$-0.423$, 0.419]). Conditional predicted attempts were 2.29 with prior exposure and 2.26 without; predicted times were approximately 19.1 seconds in both groups. These elapsed times begin at the first submission, so they do not include the work needed to create the initial solution.

\textbf{Student and exercise variation}. The solve-rate student random-intercept SD exceeded the exercise SD (2.641 vs.\ 1.595 on the logit scale). For attempts and time, the exercise SD was larger.

\textbf{Overall}. Prior same-part exposure was associated with lower final-project solve probability, while associations with attempts through first success and elapsed time among solved pairs were uncertain.

\input{table-rq3-summary}

%% file: table-rq3-summary.tex
\begin{table*}[t]
\centering
\small
\caption{RQ3: Conditional predictions and prior-feedback coefficients for final-project performance, after excluding directly feedback-bearing final pairs. Cells report estimates [intervals]: 94\% HDIs for solving and attempts, and 95\% CIs for time. Predictions set student/exercise random effects to zero; time predictions are conditional medians. Solving uses 2,242 pairs; attempts/time use 2,127 solved pairs completed within 72 hours. Coefficients compare prior feedback with no prior feedback on the log-odds, log-count, and $\log(1+\mathrm{seconds})$ scales, respectively.}
\label{tab:rq3-summary}
\begin{tabular}{lrrr}
\toprule
Outcome & No prior feedback & Prior feedback & Exposure coefficient \\
\midrule
Solve probability (\%) & 99.73 [99.42, 99.96] & 99.22 [98.32, 99.89] & $-1.096$ [$-2.074$, $-0.136$] \\
Attempts & 2.26 [1.90, 2.62] & 2.29 [1.90, 2.67] & $0.013$ [$-0.099$, $0.116$] \\
Time (seconds) & 19.1 [10.6, 34.0] & 19.1 [10.0, 35.6] & $-0.002$ [$-0.423$, $0.419$] \\
\bottomrule
\end{tabular}
\end{table*}

%% file: rq4.tex
\subsection{RQ4: Performance Change After Feedback Exposure}

\textbf{Between-group comparison}. On standard exercises, improvement differed between exposed and never-exposed students for both attempts ($p<0.001$) and time ($p=0.004$). Among the 61 exposed students, median individual improvement was 0.5 attempts and 21.7 seconds; among 54 never-exposed students, it was 0.0 attempts and $-23.2$ seconds. On final projects, improvement did not differ significantly between the 53 exposed and 54 never-exposed students (attempts: $p=0.916$; time: $p=0.877$). Changes are calculated from per-student medians.

\begin{figure*}
    \centering
    \includegraphics[width=0.95\linewidth]{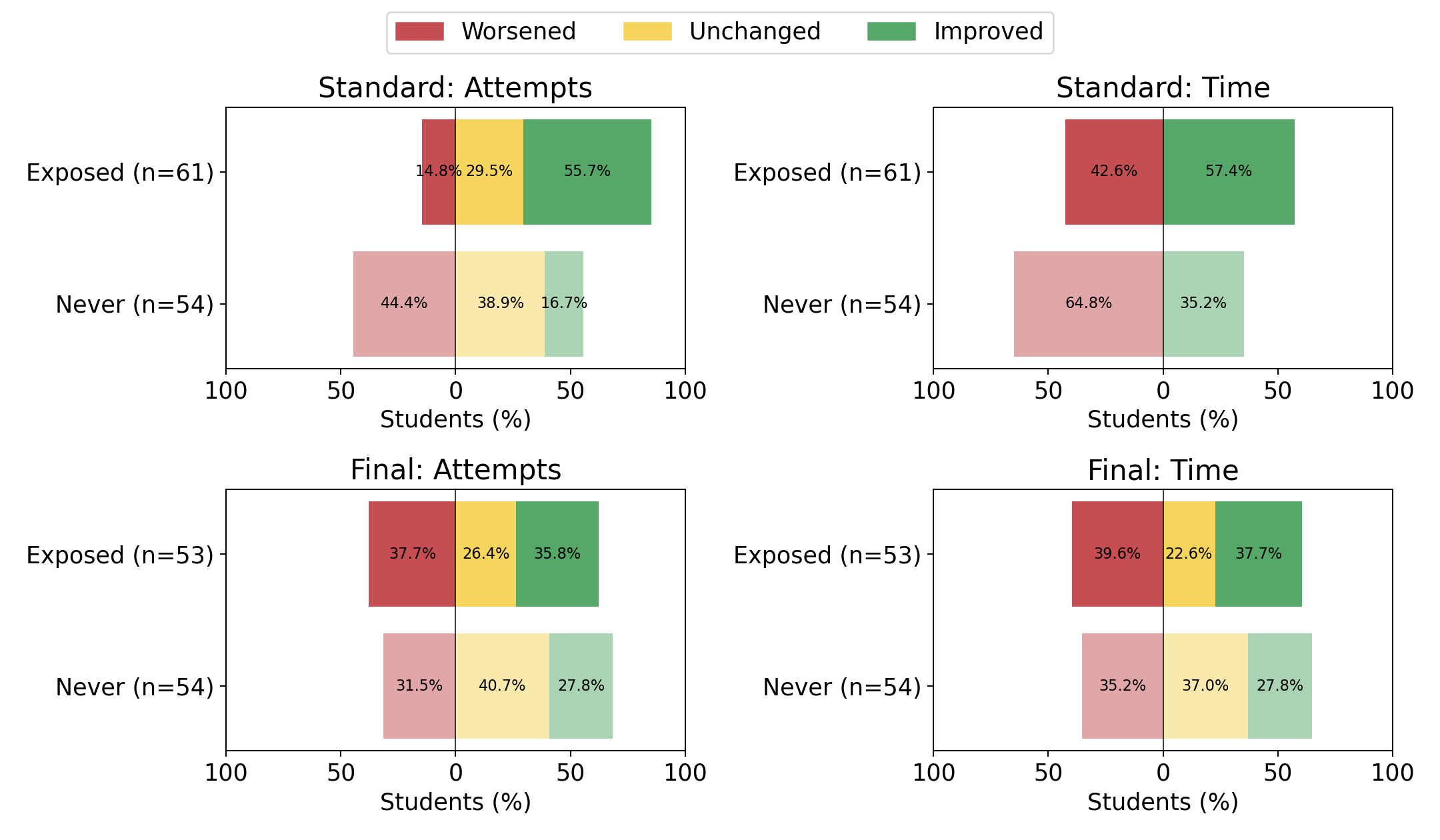}
    \caption{RQ4: Change in each student's median attempts and elapsed time to first success, using solved standard and final exercises. Exercises are assigned by their first submission: before or after first standard-exercise feedback for exposed students, or around the sequence midpoint for never-exposed students. Episodes begun before exposure can continue afterward. Lower attempts/time count as improvement. The displayed sample sizes are students. Bars use unrounded percentages; labels are rounded independently to one decimal place and may not sum to 100\%.}
    \Description{Four panels show percentages of students whose median attempts or elapsed time improved, stayed unchanged, or worsened, for standard and final exercises. Worsening extends left of zero.}
    \label{fig:rq4-improvement-after-exposure}
\end{figure*}

\textbf{Within-group comparison}. The median alone does not capture the full picture (Figure~\ref{fig:rq4-improvement-after-exposure}). On standard exercises, 55.7\% of exposed students needed fewer attempts on exercises begun after exposure, while 14.8\% needed more. The signed-rank test indicated a shift toward fewer attempts ($p<0.001$); the group median fell from 3.0 to 2.5. Never-exposed students showed changes in the opposite direction: 44.4\% needed more attempts, and 64.8\% took longer in the second half of their sequence (attempts: $p=0.042$; time: $p=0.020$). The exposed group's time change did not reach the exploratory threshold ($p=0.059$). Neither group showed a significant before/after change on final projects. For exposed students, the final-project group median nevertheless rose from 1.5 to 2.0 attempts, while the median individual change was zero; these summaries need not move together.

\textbf{Final-project steps begun before and after exposure}. The complementary timing comparison uses one median per student in each timing group. Among final-project steps begun before first feedback exposure, eventual feedback recipients had higher student-level median attempts and elapsed times than never-exposed students (attempts: 1.5 vs.\ 1.0, $p=0.003$; time: 36.8s vs.\ 0.0s, $p=0.004$). The corresponding comparisons for steps begun after exposure also differed ($p<0.001$ for both outcomes). Among the 53 students represented in both groups, neither paired comparison was significant (attempts: $p=0.720$; time: $p=0.693$). This analysis concerns attempts and elapsed time among solved pairs, with a student-wide exposure definition; it does not directly explain RQ3's part-local solve-rate association.

\textbf{Interpretation}. Standard-exercise performance changed more favorably in the exposed group, especially for attempts. Within-student changes on final projects remained uncertain. The different course stages and exercises represented on either side of the split can contribute to these patterns. All tests are exploratory, without multiplicity adjustment.

%% file: 50-discussion.tex
\section{Discussion}

\textbf{Feedback is associated with more favorable immediate outcomes, but not consistently.} Receiving feedback was associated with higher next-submission correctness in several format-period comparisons (RQ1a) and fewer attempts through first success in Periods 1 and 2 among quickly solved pairs (RQ1b). Associations with elapsed time varied: both formats were associated with faster solving in Period 1, while Diagnostic Feedback was associated with slower solving in Period 2. Model version changed alongside calendar period, cohorts, and the opt-in policy, so these differences cannot be assigned solely to the model.

\textbf{Longer feedback is associated with less favorable subsequent performance.} In RQ2, longer feedback was associated with lower next-submission correctness, more remaining attempts, and longer time to the next submission. Complexity had no clear association with correctness or attempts and a smaller positive association with time. Format showed a qualified trade-off: Diagnostic Feedback was associated with fewer remaining attempts in all three periods, but longer resubmission time was supported only in Periods 1 and 2. These results do not establish that shortening otherwise equivalent feedback would improve performance.

\textbf{Student and exercise differences depend on the outcome and population.} Exercise random-intercept SDs exceeded student SDs in the next-submission and feedback-characteristic models. Student SDs were larger for eventual solving in RQ1b and final-project solving in RQ3. These SDs describe heterogeneity on each model's link scale. The models do not identify the processes underlying that heterogeneity. Qualitative work shows that novices can complete tasks successfully with GenAI while engaging in markedly different metacognitive processes \cite{prather2024widening}. This reinforces the distinction between observed task performance and understanding, rather than explaining the random-effect differences in the present data.

\textbf{Final-project differences do not establish a causal explanation.} Prior same-part feedback was associated with lower final-project solve probability (RQ3), while associations with attempts and time among solved pairs remained uncertain. The complementary student-level timing comparison (RQ4) found differences in attempts and time on final-project steps begun before first exposure and no clear paired change across steps begun before and after exposure. Selection remains a plausible explanation, but these comparisons do not identify that mechanism or isolate performance completed before exposure. The two analyses also differ in exposure definition, outcome, and time restrictions; RQ4 cannot directly explain RQ3's solve-rate coefficient.

\textbf{Individual changes and group medians tell different stories.} Standard-exercise improvement differed between exposed and never-exposed students. Among exposed students, 55.7\% needed fewer attempts on exercises begun after exposure and 14.8\% needed more, while never-exposed students more often worsened. Final-project improvement comparisons remained uncertain. A zero median individual change does not mean group medians stayed fixed: exposed students' final-project median attempts increased from 1.5 to 2.0. The different course-stage splits and unadjusted exploratory tests limit interpretation.

\textbf{Practical implication.} Bounded feedback length and a choice between a single-issue hint and a fuller diagnostic account are plausible design directions. However, students' preferences need not identify the format associated with better performance: Santos et al. found that participants preferred baseline GPT-4o explanations despite resolving errors faster with fine-tuned explanations \cite{11176024}. Evaluations of student choice should therefore measure preference, performance, and understanding separately. The present associations motivate those evaluations rather than demonstrating that either design improves learning.

%% file: 55-limitations.tex
\section{Limitations}
This study has several limitations.
First, model version and calendar period coincide: each deployment used a different underlying model, so period differences can also reflect cohort, calendar, opt-in policy, or unmeasured course changes. Prompts, generation settings, and aspects of ordinary grader feedback also changed during the study. Single-Issue and Diagnostic formats occur within all three periods; feedback format itself is not fully confounded with period.
Second, the analyzed no-feedback groups include submissions outside active feedback availability and are not pure randomized controls. Feedback exposure was not randomized in Periods~2 and~3. Opting in enabled automatic feedback after eligible incorrect submissions, so messages could appear even when students already understood the error or could correct it without assistance. We did not measure whether each message was needed, read, or used. Prior difficulty, motivation, and feedback-seeking can therefore confound exposure and content associations. The length analysis estimates an association, not the effect of experimentally shortening an otherwise equivalent message. RQ1b classifies messages received during an exercise episode, so exposure can itself depend on the number of attempts and time spent struggling.
Third, RQ3 and RQ4 use a restricted, highly engaged Period~2 cohort. Their outcome-specific samples are reported in \autoref{tab:experiment_ns}; they do not represent all enrolled students or verified course completers.
Fourth, RQ1a and RQ2 require a known next submission within 30 minutes, and RQ1b attempts/time condition on solving within that interval. RQ1b's eventual-solving model retains slow successes and observed failures. RQ3 attempts/time analyses exclude pairs taking more than 72 hours to reach first success, whereas the descriptive RQ4 comparisons retain longer gaps. Elapsed submission gaps include grading and AI-generation delays as well as student activity; they do not isolate time spent reading feedback or correcting code. A first-attempt success has zero logged time to solve.
Fifth, eventual solving is common, limiting detectable changes in that outcome. Predicted values with random effects set to zero are conditional model summaries, not observed group rates or averages integrated over the student/exercise population.
Sixth, RQ4 assigns exercise episodes by their first submission; some episodes begun before feedback exposure continued afterward. The before group therefore does not isolate performance completed before exposure. Its exposure and midpoint splits also need not occur at comparable course stages. Its complementary timing comparison uses student-level medians and paired tests where appropriate, but does not control all differences in the projects attempted. Tests are exploratory and unadjusted for multiplicity; non-significance does not establish unchanged performance or causal absence.
Finally, this single-course study measures behavioral performance---correctness, attempts, and elapsed time---rather than direct learning or understanding. The temporal final-project analysis is a step toward follow-up performance, but learning gains and the effects of bounded length or student-selected feedback formats require further evaluation.

%% file: 60-conclusions.tex
\section{Conclusions and Future Work}

We studied three iterations of an LLM-generated feedback system deployed in a live programming course, across three models and three feedback formats. Feedback was associated with higher next-submission correctness and fewer attempts through first success in several comparisons, but associations varied by period and outcome. Longer messages were associated with lower next-submission correctness, more remaining attempts, and longer resubmission time in pooled models. Prior same-part exposure was associated with lower final-project solve probability. Student-level comparisons showed differences on final-project steps begun before exposure and uncertain within-student changes; they do not identify the cause of those differences. Standard-exercise changes were more favorable in the exposed group, but the exposure and midpoint splits need not represent comparable course stages.

These results suggest two directions to test. First, feedback systems could place bounds on response length while preserving useful guidance. Second, courses could evaluate letting students choose their preferred format.

Future work should use randomized feedback availability and message-design comparisons, investigate pre-existing differences between feedback-exposed and never-exposed students, and move beyond performance measures toward direct evaluation of learning and long-term understanding.